\documentclass[twocolumn,amsmath,amssymb,superscriptaddress]{revtex4}

\usepackage{graphicx}
\usepackage{dcolumn}
\usepackage{bm}
\usepackage{color}
\def\Vec#1{\bm{#1}}

\begin{document}


\title{
N-independent Localized Krylov Bogoliubov-de Gennes Method: \\
Ultra-fast Numerical Approach to Large-scale Inhomogeneous Superconductors}
%

\author{Yuki Nagai}
\email[Corresponding author: ]{nagai.yuki@jaea.go.jp}
\affiliation{CCSE, Japan  Atomic Energy Agency, 178-4-4, Wakashiba, Kashiwa, Chiba, 277-0871, Japan}
\affiliation{
Mathematical Science Team, RIKEN Center for Advanced Intelligence Project (AIP), 1-4-1 Nihonbashi, Chuo-ku, Tokyo 103-0027, Japan
}

\date{\today}

\begin{abstract}
We propose the ultra-fast numerical approach to large-scale inhomogeneous superconductors, which we call the Localized Krylov Bogoliubov-de Gennes method (LK-BdG). 
In the LK-BdG method, the computational complexity of the local Green's function, which is used to calculate the local density of states and the mean-fields, does {\it not} depend on the system size $N$. 
The calculation cost of self-consistent calculations is ${\cal O}(N)$, which enables us to open a new avenue for treating extremely large systems with millions of lattice sites. 
To show the power of the LK-BdG method, we demonstrate a self-consistent calculation on the 143806-site Penrose quasicrystal lattice with a vortex and a calculation on 1016064-site two-dimensional nearest-neighbor square-lattice tight-binding model with many vortices.
We also demonstrate that it takes less than 30 seconds with one CPU core to calculate the local density of states with whole energy range in 100-millions-site tight-binding model.

\end{abstract}

\maketitle

\section{Introduction}
The mean-field approach through the Bogoliubov-de Gennes (BdG) equations is one of the most convenient and efficient ways to describe inhomogeneous superconductivity. 
In the past two decades, studies about quasiparticle excitations in superconducting systems 
with junctions or vortices become more important since these systems can be stages for a topological quantum computing with the use of the Majorana quasiparticles\cite{Nayak2008,Teo2010,Alicea2011}.
Recently, the Majorana zero modes have been observed in many systems such as the iron-based superconductor FeTe$_x$Se$_{1-x}$
\cite{Wang2018,Machida2019}.
Since in the systems with junctions or vortices one needs to use a real-space formulation, the numerical simulation becomes computationally involved. 
Although there are alternative approaches to inhomogeneous superconductivity like quasiclassical Eilenberger theory or Ginzburg-Landau methods, these methods can not treat discretized quantum modes like Majorana zero modes. 
In addition to topological materials, there are many interesting systems to be solved such as high-$T_c$ superconductors for which the superconducting coherence length is of the order of the Fermi wavelength, or nanoscale superconductivity for which superconducting coherence length is comparable to the system size. 
Therefore, the need for a fully quantum-mechanical approach has become imperative. 

It is very hard to diagonalize the Hamiltonian in large inhomogeneous systems, since the computational complexity to diagonalize the Hamiltonian matrix is ${\cal O}(N^3)$.
Here, $N$ is the matrix size. 
In the last decade, various kinds of numerical approaches to solve the BdG equations for inhomogeneous systems have been developed\cite{Covaci2010,Nagai2012,Nagai2013,Nagai2017}.
The computational complexities of these approaches are ${\cal O}(N^2)$ for self-consistent calculations and ${\cal O}(N)$ for calculating the local quantities like the local density of states (LDOS), respectively.
However, if one wants to treat inhomogeneous systems with internal degrees of freedom ({\it e.g.} the iron-based superconductors are multi-band systems), the computational complexity becomes huge even with the use of supercomputing systems with thousands of CPU cores. 
Therefore, it is still hard to treat large realistic inhomogeneous systems. 

In this paper, by focusing on the fact that the one-particle local Green's function is constructed locally in real space, we propose the ultra-fast numerical approach to large-scale inhomogeneous superconductors, which we call the Localized Krylov Bogoliubov-de Gennes method (LK-BdG). 
We show that vectors in the Krylov subspace to calculate the Green's function are localized. 
The computational complexities in the LK-BdG are ${\cal O}(N)$ for self-consistent calculations and ${\cal O}(1)$ for calculating the local quantities, respectively.  
To show the power of the LK-BdG method, we demonstrate a self-consistent calculation on the $s$-wave superconducting 143806-site Penrose quasicrystal lattice with a vortex and a calculation on 1016064-site two-dimensional $s$-wave nearest-neighbor square-lattice tight-binding model with many vortices. 
Finally, the summary is given.

\section{Method}
\subsection{Model}
The Bogoliubov-de Gennes equations describe the behavior of electrons and holes in superconductors, 
which are coupled to mean-fields. 
A general BdG Hamiltonian is given as ${\cal H} = \Psi^{\dagger} \hat{H} \Psi/2$. 
The column vector $\Psi$ is composed of $N$ fermionic annihilation $c_i$ and creation operators 
$c_i^{\dagger}$ ($i = 1,2,\cdots,N$), $\Psi = (\{ c_i \},\{ c_i^{\dagger} \} )^{\rm T}$, 
where $\{ c_i \} = (c_1,c_2,\cdots,c_N)^{\rm T}$ and 
$\{c_i^{\dagger} \} = (c_1^{\dagger},c_2^{\dagger},\cdots,c_N^{\dagger})$. 
Here, $N$ is the size of the normal state Hamiltonian matrix. 
The subscription $i$ in $c_i$ or $c_i^{\dagger}$ indicates a quantum index depending on spatial site, spin, orbital, etc. 
For example, in a single-band spin-singlet superconductor with $N$ spatial sites, we use $\{ c_{i\uparrow} \} = (c_{1\uparrow},c_{2\uparrow},\cdots,c_{N\uparrow})^{\rm T}$ and 
$\{c_{i \downarrow}^{\dagger} \} = (c_{1\downarrow}^{\dagger},c_{2\downarrow}^{\dagger},\cdots,c_{N\downarrow}^{\dagger})$.
The Hamiltonian matrix $\hat{H}$ is a $2N \times 2N$ Hermitian matrix given as  
\begin{align}
    {\hat H} = \left( \begin{matrix}
        \hat{H}^{\rm N} & \hat{\Delta} \\
        \hat{\Delta}^{\dagger} & - \hat{H}^{\rm N *}
    \end{matrix} \right).
\end{align}
Here, $\hat{H}^{\rm N}$ is a Hamiltonian matrix in normal states and $\hat{\Delta}$ is a 
superconducting order parameter.

Without diagonalizing the BdG Hamiltonian directly, we can calculate physical observables and mean-fields 
with the use of the one-particle Green's function $\hat{G}(z) = (z \hat{I} - {\hat H})^{-1}$. 
The important quantity is the difference of the retarded and advanced Green's function matrices determined as 
$\hat{d}(\omega) = \hat{G}^{\rm R}(\omega) - \hat{G}^{\rm A}(\omega)$. 
For example, the LDOS with a quantum index $i$ and the mean-field $\langle c_{i} c_{j} \rangle$ are respectively expressed as\cite{Covaci2010, Nagai2012} 
\begin{align}
N(\omega, i) &= -\frac{1}{2 \pi i} {\bm e}(i)^{\rm T} \hat{d}(\omega) {\bm e}(i), \\
\langle c_{i} c_{j}  \rangle &= \frac{1}{2 \pi i} \int_{-\infty}^{\infty} d\omega {\bm e}(j)^{\rm T} \hat{d}(\omega) {\bm h}(i).
\end{align}
Here, ${\bm e}(i)$ and ${\bm h}(i)$ are $2N$-component unit-vector defined as 
\begin{align}
    [{\bm e}(i)]_{\gamma} = \delta_{i\gamma}, \: [{\bm h}(i)]_{\gamma} = \delta_{i+N,\gamma}. \label{eq:h}
\end{align}

\subsection{Localized Krylov subspace}
\begin{figure}[t]
    \begin{center}
         \begin{tabular}{p{ 1 \columnwidth}} 
          \resizebox{1 \columnwidth}{!}{\includegraphics{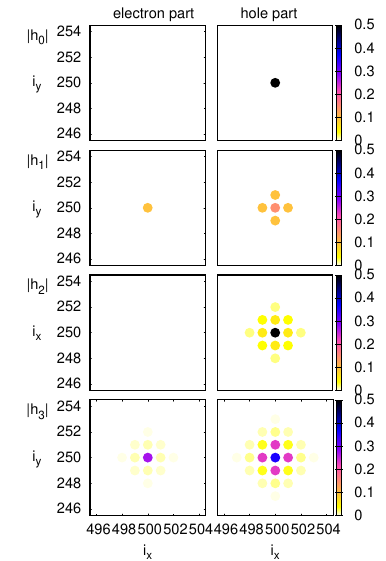}} 
        \end{tabular}
    \end{center}
    \caption{(Color online)
    Spatial dependence of the amplitude of the Chebyshev vector $|{\bm h}_m(i)|$ in the two-dimensional $1000 \times 1000$ square-lattice nearest-neighbor tight-binding model with $s$-wave order parameter and a vortex located at a center, where the site index $i$ is $i = (500,250)$. 
    Left panels: the electron part of the vector. Right panels: the hole part of the vector. 
    \label{fig:fig1}
     }
    \end{figure}

We focus on the fact that the vectors ${\bm e}(i)$ and ${\bm h}(i)$ are localized in real space.
We introduce the order-$m$ Krylov subspace generated by the Hamiltonian matrix $\hat{H}$ and 
an unit vector ${\bm b}$ ($= {\bm e}(i)$ or ${\bm h}(i)$) given as 
\begin{align}
    {\cal K}_{m}(\hat{H},{\bm b}) = {\rm span} \: \{ {\bm b},\hat{H} {\bm b},\hat{H}^2 {\bm b},\cdots,\hat{H}^{m-1} {\bm b} \}.
\end{align}
We call ${\cal K}_{m}(\hat{H},{\bm b})$ the {\it localized} Krylov subspace, since $m$ vectors are localized as follows. 
The element of the second vector is expressed as 
\begin{align}
    [\hat{H} {\bm e}(i)]_k &= \sum_{l=1}^{2N} \hat{H}_{kl} [{\bm e}(i)]_l = \hat{H}_{ki}.
\end{align}
If the Hamiltonian matrix $\hat{H}$ is sparse, the number of finite value elements of  
$[\hat{H} {\bm e}(i)]_k$ is a few. 
For example, in the case of the two-dimensional square-lattice tight-binding model with nearest neighbor hopping, 
there are only four elements in the normal state Hamiltonian matrix. 
The element of the third vector is expressed as 
    $[\hat{H}^2 {\bm e}(i)]_k = \sum_{l} \hat{H}_{kl} \hat{H}_{li}$, 
where the index $l$ for the summation is restricted due to the sparseness of the Hamiltonian. 
Thus, the number of the finite value elements of $m$-th vector is $\sim m^d$, where $d$ is the dimension of the system.  
With the use of the above discussion, we can reduce the computational complexity of the matrix-vector product: 
\begin{align}
[\hat{H} {\bm v}]_k &= \sum_{\{ l  | |v_{l}| > 0 \} } \hat{H}_{kl} v_l, 
\end{align}
where the number of the indices in the summation does {\it not} depend on the size of the system $N$, as shown in Fig.~\ref{fig:fig1}.
The first vector $\Vec{h}_0$, defined in Eq.~(\ref{eq:h}), is localized in hole-space. 
Although the amplitude of the vectors spreads in electron- and hole- spaces due to the superconducting order parameter matrix $\hat{\Delta}$, the amplitudes of $\Vec{h}_1$, $\Vec{h}_2$, $\Vec{h}_3$ are still localized (See, Fig.~\ref{fig:fig1}). 
We should note that this property is useless for finding eigenvalues and eigenvectors 
with the use of the Krylov-subspace based methods such as Lanczos or Arnoldi methods, 
since the eigenvectors are usually not localized so that we should take large $m$ in the Krylov subspace where the elements of the $m$-th vector are all finite. 

\subsection{Fast matrix-vector product}
We describe the fast matrix-vector product in details. 
An element of the matrix-vector product $\hat{H} \Vec{v}^k$ is expressed as 
\begin{align}
\left[ \hat{H} \Vec{v}^k \right]_i &= \sum_l H_{il} v_l^k.
\end{align}
Here, we assume that the vector $\Vec{v}^k$ is localized around an index $k$. 
We define a subspace ${\cal N}_{\epsilon}(\Vec{v}^k)$ as a set of indices where absolute values of the elements of the vector $\Vec{v}^k$ exceed the threshold $\epsilon$:
\begin{align}
{\cal N}_{\epsilon}(\Vec{v}^k) &\equiv \{i | \: |v_i^k| > \epsilon \}.
\end{align}
The threshold $\epsilon$ has been introduced for a truncation of the matrix-vector product in Ref.~\onlinecite{Furukawa2004}, 
which accelerates BdG simulations more effectively. 
In this paper, we adopt $\epsilon = 10^{-6}$.

We can restrict the matrix-vector product operation within this subspace ${\cal N}_0(\Vec{v}^k)$ since 
\begin{align}
|v_i^k| \le \epsilon \: \: \: {\rm if} \: \: i \notin {\cal N}_{\epsilon}(\Vec{v}^k).
\end{align}
The number of large elements ($> \epsilon$) of the vector $\hat{H} \Vec{v}^k$ increases after the matrix-vector product, which depends on the sparseness of the hermitian matrix $\hat{H}$.  
We define a subspace ${\cal M}(i,\hat{H})$ as a set of indices of non-zero elements on the $i$-th row: 
\begin{align}
{\cal M}(i,\hat{H}) &\equiv \{j | \: H_{ij} \neq 0 \}.
\end{align}
Thus, the $i$-th element of the vector $[\Vec{v}^k]_i$ contributes to $j (\in {\cal M}(i,\hat{H}))$-th element of the vector $[\hat{H} \Vec{v}^k]_j$. 
With the use of these properties, the algorithm is given in Table.~\ref{table:Hx}.  
In the Table.~\ref{table:Hx}, we use the hermitian property of the matrix $\hat{H}$. 
In the actual code, we prepare $N$-dimensional arrays $\Vec{v}^k$, ${\cal N}_{\epsilon}(\Vec{v}^k)$ and a scalar $n$, where $n$ is the number of finite elements of $\Vec{v}^k$. 
Here, we store the index of the $i$-th finite element of the vector $\Vec{v}^k$ in $i$-th element of ${\cal N}_{\epsilon}(\Vec{v}^k)$.

\begin{table}[tbh]
  \caption{Algorithm of the fast matrix-vector product $\Vec{v}_{\rm out}^k \equiv \hat{H} \Vec{v}^k$. Here, the vector $\Vec{v}^k$ keeps non-zero element determined by ${\cal N}_{\epsilon}(\Vec{v}^k)$ only. 
  sets of indices ${\cal N}_{\epsilon}(\Vec{v}_{\rm temp})$ and ${\cal N}_{\epsilon}(\Vec{v}_{\rm out}^k)$ should be updated in the loops. 
  }
  \label{table:Hx}
  \begin{center}
    \begin{tabular}{ c l }
      \hline 
      1. & Input $\Vec{v}^k$, ${\cal N}_{\epsilon}(\Vec{v}^k)$, $\hat{H}$, and a set of ${\cal M}(i,\hat{H})$\\
      2. &Output $\Vec{v}_{\rm out}^k \equiv \hat{H} \Vec{v}^k$ \\
      3. & Set the temporal vector $\Vec{v}_{\rm temp}=0$ \\
      4.  & For $i \in {\cal N}_{\epsilon}(\Vec{v}^k)$  Do: \\
      5  & ~~~~~~For $j \in {\cal M}(i,\hat{H})$ Do: \\
      6.  & ~~~~~~~~~~~~$[\Vec{v}_{\rm temp}]_j = [\Vec{v}_{\rm temp}]_j  + H_{ij}^{\ast} [\Vec{v}^k]_i$ \\
      7. & ~~~~~~End Do \\
      8. & End Do \\
      9.  & For $i \in {\cal N}_{\epsilon}(\Vec{v}_{\rm temp})$  Do: \\
      10. & ~~~~~~$[\Vec{v}_{\rm out}^k]_i = [\Vec{v}_{\rm temp}]_i$ \\
      11. & End Do \\
    \hline 
    \end{tabular}
  \end{center}
\end{table}

\subsection{Chebyshev polynomial method}
In the Chebyshev polynomial method, we generate vectors by a recurrence formula: 
\begin{align}
    {\bm q}_{n+1} &= 2 \hat{K} {\bm q}_n - {\bm q}_{n-1} \: (n \ge 2), 
\end{align}
with ${\bm q}_0 = {\bm b}$, ${\bm q}_1 = \hat{K} {\bm b}$ and $\hat{K} = (\hat{H}-b \hat{I})/a$\cite{Covaci2010,Nagai2012}.
$n$ generated vectors are in the Krylov subspace ${\cal K}_{n+1}(\hat{H},{\bm b})$. 
The mean-field $\langle c_i c_j \rangle$ is expressed as $\langle c_i c_j \rangle = \sum_{n=0}^{\infty} {\bm e}(j)^{\rm T} {\bm h}_n(i) {\cal T}_n/w_n$ where ${\cal T}_n = \int_{-1}^{1}dx f(ax+b)W(x)\phi_n(x)$, $W(x)= 1/\sqrt(1-x^2)$, 
$w_n = (1+\delta_{n0})\pi/2$, $\phi_n(x) = \cos (n \arccos(x))$ and $f(x) = 1/(1+\exp(x/T))$.  
We found that the mean-field $\langle c_i c_j \rangle$ can be calculated with good enough accuracy in the localized Krylov subspace whose order $m$ is much smaller than the dimension of the matrix if the index $j$ is not far from the index $i$ in real space. 
Therefore, the Chebyshev polynomial method with the localized Krylov subspace to calculate elements of the matrix $\hat{d}(\omega)$ does not depend on the matrix dimension $N$. 
We note that Furukawa and Motome have shown that the free energy in the Monte Carlo simulations can be calculated with the use of the similar localized Krylov subspace in fermion systems coupled with classical degree of freedom\cite{Furukawa2004}.

\subsection{Lanczos method for a Green's function.}
Another Krylov-subspace-based method to calculate a Green's function is the Lanczos method\cite{Dagotto1994}. 
The diagonal elements of Green's function $[\hat{G}(z)]_{ii}$ $(i \le N)$ can be calculated by the continued fraction expansion expressed as $[\hat{G}(z)]_{ii} = 1/(z- a_0 -b_1^2/(z-a_1 -b_2^2/(z-a_2 - \cdots)))$, where the coefficients $a_n$ and $b_n^2$ are calculated by a recurrence formula: 
\begin{align}
{\bm j}_{n+1} &= \hat{H} {\bm j}_{n} - a_n {\bm j}_n - b_n^2 {\bm j}_{n-1}, 
\end{align}
with  $a_n = {\bm j}_n^{\rm T} \hat{H} {\bm j}_n/{\bm j}_n^{\rm T}{\bm j}_n$ 
and $b_n^2 = {\bm j}_n^{\rm T} {\bm j}_n/{\bm j}_{n-1}^{\rm T} {\bm j}_{n-1}$, 
supplemented by $b_0^2 = 0$, ${\bm j}_{-1} = 0$ and ${\bm j}_{0} = {\bm e}(i)$. 
In the Lanczos method for a Green's function, the $n$-th order Krylov subspace is given as ${\cal K}_n(\hat{H},{\bm e}(i))$, which is localized. 
We note that the Lanczos method with the localized Krylov subspace has been used in the field of the order-$N$ first-principles calculations\cite{Ozaki2000,Ozaki2006}.
We can not use the Lanczos method to calculate the superconducting mean-field $\langle c_i c_j \rangle$, since the mean-field is calculated from the off-diagonal element of the Green's function. 
When we calculate the LDOS, the Lanczos method becomes one of the choices. 
We numerically find that the energy resolution of the LDOS obtained by the Lanczos method is better than that obtained by the Chebyshev polynomial method if the order of the Krylov subspace is same. 
Therefore, in this paper, we adopt the Lanczos method to calculate the LDOS. 

\subsection{Computational complexity}

We summarize computational complexities of different methods as shown in Table \ref{table:2}. 
The computational complexity for the local quantity such as the local density of states is ${\cal O}(1)$ in our method, since the computational complexity of the matrix vector product is ${\cal O}(1)$.
This complexity is proportional to how many non-zero elements exist in the vector $\Vec{v}^k$, which depends on dimension of systems. 
In a $d$-dimensional system, the number of the non-zero elements in the vector is proportional to $m^d$. Here, $m$ is the number of matrix-vector products. 
Since the number of the non-zero element is smaller than the number of all elements, this method is always faster than the other Krylov based method.

\begin{table}[t]
\caption{Computational complexities. Here, $N$ is the matrix size of the Hamiltonian. }
\label{table:2}
\begin{ruledtabular}
\begin{tabular}{cccccc}
Algorithms &
Local quantity  & self-consistent calculation \\
Full diagonalization & ${\cal O}(N^3)$ &  ${\cal O}(N^3)$ \\
Krylov \cite{Covaci2010,Nagai2012,Nagai2013,Nagai2017} & ${\cal O}(N)$ & ${\cal O}(N^2)$\\
Localized Krylov & ${\cal O}(1)$ & ${\cal O}(N)$ \\
\end{tabular}
\end{ruledtabular} 
\end{table}

\section{Demonstrations}
\subsection{Demonstration I: quasicrystal with the Penrose lattice}
\begin{figure}[t]
        \begin{center}
             \begin{tabular}{p{ 1 \columnwidth}} 
              \resizebox{1 \columnwidth}{!}{\includegraphics{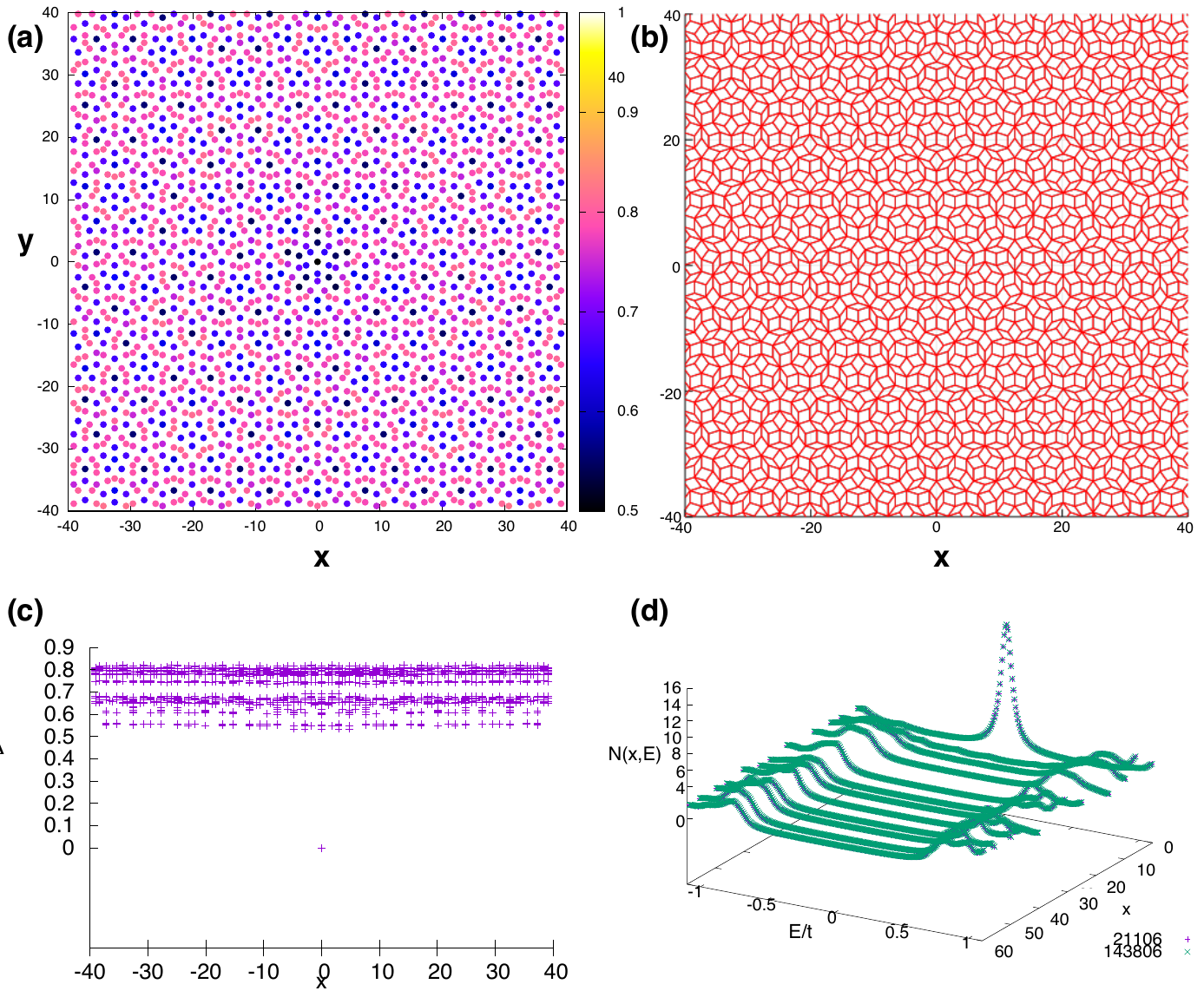}} 
            \end{tabular}
        \end{center}
        \caption{(Color online)
        (a) Gap amplitude around a vortex on the 143806-site Penrose lattice. (b) Penrose lattice around a center. 
        (c) View from the side of a Gap amplitude plot. 
        (d) The local density of states around a vortex as a functional of the energy on $y$-axis on the 143806-site and 21106-site Penrose lattices. 
        \label{fig:pen}
         }
    \end{figure}

We demonstrate a self-consistent calculation of quasicrystalline superconductors to show the power of the LK-BdG method. 
Recently, the superconductivity of quasicrystals was discovered in Al-Zn-Mg quasicrystalline alloys\cite{Kamiya2018}. 
Sakai and Arita have studied possible superconductivity on a Penrose-tiling structure, which is a prototype of quasicrystalline structures\cite{Sakai2017,Sakai2019}. 
They found that there exists a superconducting state with spatially extended Cooper pairs in the attractive Hubbard model. 
This Penrose-tiling structure that we call Penrose lattice is one of good demonstrations for the LK-BdG, since the inhomogeneous superconducting order parameter naturally appears. 
We consider the tight-binding model on the Penrose lattice proposed in the previous papers\cite{Sakai2017,Sakai2019} (Also see the supplemental materials\cite{sup}). 
We introduce a vortex located at a center. 
We calculate a $s$-wave onsite superconducting order parameter $\langle c_i c_i \rangle$ with the interaction $U = -3t$ and 
the chemical potential $\mu = -1t$ at the zero temperature.                                                                                                                    
The Chebyshev cutoff parameter is $n_c = 200$, which is enough to obtain the mean-field.
The renormalized parameters $a$ and $b$ for the Hamiltonian matrix are $a = 10t$ and $b =0$, respectively. 
In Fig.~\ref{fig:pen}, we show the self-consistent solution on the 143806-site Penrose lattice (Also see Fig.~1 in the supplemental materials\cite{sup}). 
By comparing with 143806-site and 21106-site systems, 
we show that a center region can be regarded as a bulk since the LDOS around a center 
does not depend on the lattice size as shown in Fig.~\ref{fig:pen}(d). 
While the level spacing of vortex bound states in conventional systems is characterized by $\Delta^2/E_{\rm F}$ with the Fermi energy $E_{\rm F}$\cite{Hayashi1998}, 
the energy level in the Penrose lattice is close to zero. 
The size of the vortex core is small as shown in Fig.~\ref{fig:pen}(a).
In the conventional theory for the vortex bound states, a minimum energy level in a small vortex core is high due to a quantum confinement of quasiparticles.  
Note that this energy level depends on the position of a vortex center\cite{note}.   
Because there is no Fermi wave length due to the absence of the translational symmetry, 
this suggests that interesting vortex physics exists in quasicrystalline superconductors. 

\begin{figure}[t]
        \begin{center}
             \begin{tabular}{p{ 1 \columnwidth}} 
              \resizebox{1 \columnwidth}{!}{\includegraphics{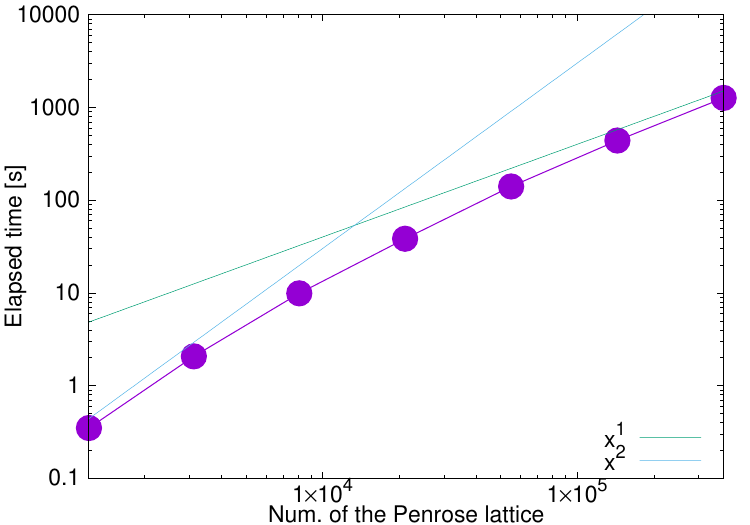}} 
            \end{tabular}
        \end{center}
        \caption{(Color online)
        Elapsed time for one self-consistent iteration step on the Penrose lattice. We use 240 CPU cores on the supercomputing system SGI ICE X at the Japan Atomic Energy Agency. 
        The number of the lattice sites in the largest system is 375971.
        \label{fig:elaps}
         }
    \end{figure}

Let us show the system size dependence of the computational complexity for self-consistent calculations. 
Figure \ref{fig:elaps} shows that the computational complexity is ${\cal O}(N)$. 
For example, the elapsed time for one iteration step in 375971-site (143806-site) Penrose lattice with 240 CPU cores on the supercomputing system SGI ICE X at the Japan Atomic Energy Agency \cite{note2} is about 1265 (440) seconds.

\subsection{Demonstration II: Ultra-large 2D tight-binding model}
\begin{figure}[t]
        \begin{center}
             \begin{tabular}{p{ 1 \columnwidth}} 
              \resizebox{1\columnwidth}{!}{\includegraphics{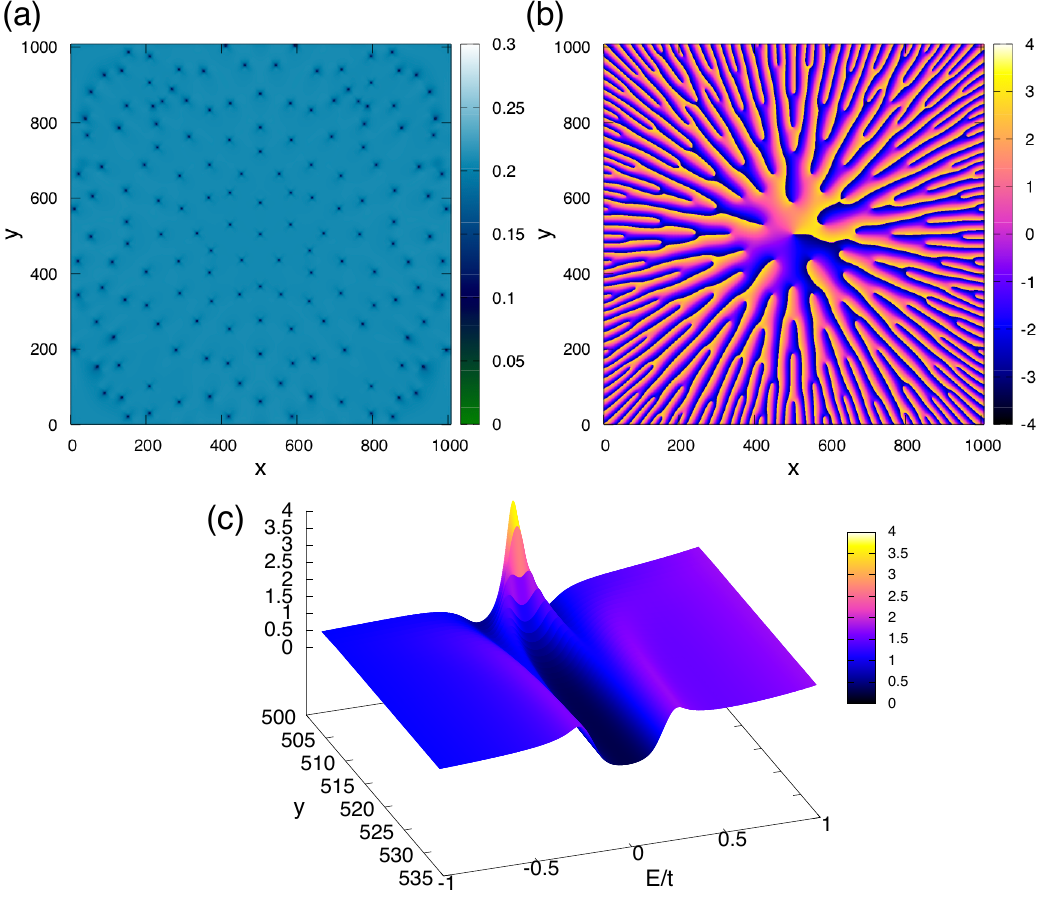}} 
            \end{tabular}
        \end{center}
        \caption{(Color online)
        (a) Gap amplitude on the 1008 $\times$ 1008 2D square lattice. 
        (b)  Phase of the gap.
        We introduce vortices with the Penrose tiling pattern to make an initial guess of the phase of the gap. 
        (c) The local density of states around a vortex with fixing $x = 508$.
        \label{fig:gap}
         }
    \end{figure}

We demonstrate that the LK-BdG method can treat ultra-large 2D tight-binding model.
 We consider 2D $N_x \times N_y$ square-lattice nearest-neighbor tight-binding model with many vortices at zero temperature.
 We calculate the $s$-wave order parameter with $U = -2.4t$, $\mu = -1.5t$, $n_{c} = 200$, $a = 10t$, $b=0$.
 We consider the Peierls phase to introduce vortices perpendicular to the system\cite{Smith2016,Nagai2012}. 
 The symmetric gauge $\Vec{A}({\bm r}) = (1/2) {\bm H} \times {\bm r}$ with ${\bm H} = (0,0,H_z)$ is used.
 Here, we consider $m$ vortices in the system ($H_z = m \phi_0/(N_x N_y)$) and the type II limit (the magnetic penetration depth $\lambda \rightarrow \infty$). 
 As the initial guess of the superconducting order parameter, we introduce vortices with the Penrose tiling pattern, whose phase singularities are located at vertices of the Penrose tiling. 
Although this vortex configuration is not a true ground state, we can obtain a similar vortex configuration as a metastable state if the vortex-vortex distance is long enough in the type II-limit superconductor.
After 120 iteration steps with solving gap equations\cite{note3}, we obtain the superconducting gap distribution in the $1008 \times 1008$ 2D tight-binding model.
As shown in Fig.~\ref{fig:gap}, the LK-BdG method can calculate superconducting mean-fields and the LDOS in large systems with many vortices.

\begin{figure}[t]
        \begin{center}
             \begin{tabular}{p{ 1 \columnwidth}} 
              \resizebox{1 \columnwidth}{!}{\includegraphics{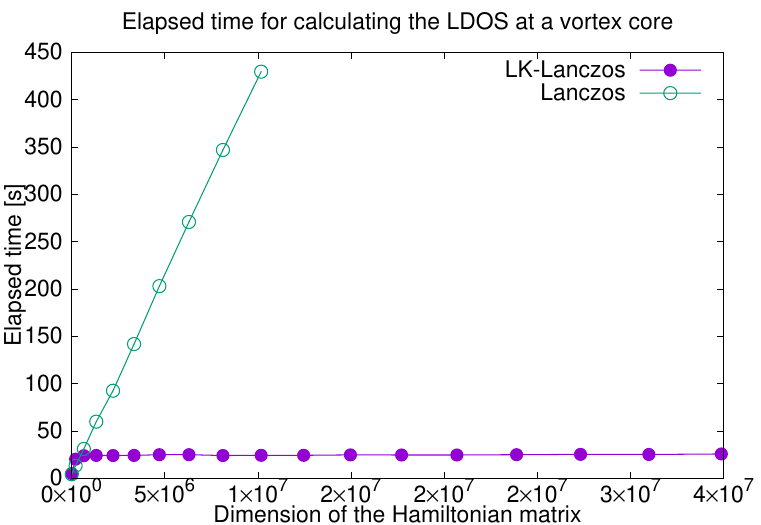}} 
            \end{tabular}
        \end{center}
        \caption{(Color online)
        Elapsed time for calculating the LDOS at a vortex core. 
        The calculations with a single CPU core are done on a laptop PC.
        \label{fig:fig6}
         }
    \end{figure}

To compare with a previous method, we measure the elapsed time for calculating the LDOS with a single vortex located at a center. 
For simplicity, we consider a single vortex whose coherence length is 10 (the unit is the lattice spacing). 
We consider 4000 energy meshes from $-2.5t$ to $5.5t$.  
The number of the Lanczos iterations is 400 and the smearing factor for the LDOS is $5 \times 10^{-2}$. 
The calculations with a single CPU core are done on a laptop PC (MacBook Pro (13-inch, 2018, Four Thunderbolt 3 ports) with 2.7GHz Intel Core i7 CPU with 4 cores). 
As shown in Fig.~\ref{fig:fig6}, the computational complexity of our method with the localized Krylov subspace does not depend on $N$, where $N$ is the dimension of the Hamiltonian matrix. 
We confirm that the elapsed time in the 5000 $\times$ 5000 lattice system, whose matrix dimension is $5 \times 10^{7}$, is only about 26 seconds on the laptop PC. 
On the other hand,  the complexity of the previous Lanczos-based method is ${\cal O}(N)$ as shown in Fig.~
\ref{fig:fig6}.

\section{Summary}
We proposed the LK-BdG, the ultra-fast numerical approach to large-scale inhomogeneous superconductors, by focusing on the fact that the vectors in the Krylov subspace for the Green's function is localized.  
In a self-consistent calculation, the computational complexity is ${\cal O}(N)$. 
We also showed that the computational complexity to calculate the local density of states does not depend on the system size $N$. 
The LK-BdG method enables us to open a new avenue for treating extremely large systems with millions of lattice sites.

\section*{Acknowledgment}
We would like to thank S. Yamada and M. Machida for helpful discussions.
The calculations were performed by the supercomputing system SGI ICE X at the Japan Atomic Energy Agency.
This work was partially supported by JSPS-KAKENHI Grant Numbers 18K11345 and the “Topological Materials Science” (No. 18H04228) JSPS-KAKENHI on Innovative Areas.

\appendix

\clearpage
\newpage
\widetext
\onecolumngrid

\setcounter{equation}{0}
\renewcommand{\thefigure}{S\arabic{figure}} 

\setcounter{figure}{0}

\renewcommand{\thesection}{S\arabic{section}.} 
\renewcommand{\theequation}{S\arabic{equation}} 
\renewcommand{\thetable}{S\arabic{table}} 
\begin{flushleft} 
{\Large {\bf Supplemental material}}
\end{flushleft} 

\section{Penrose Tiling}
\begin{figure}[b]
    \begin{center}
         \begin{tabular}{p{ 0.7 \columnwidth}} 
          \resizebox{0.7 \columnwidth}{!}{\includegraphics{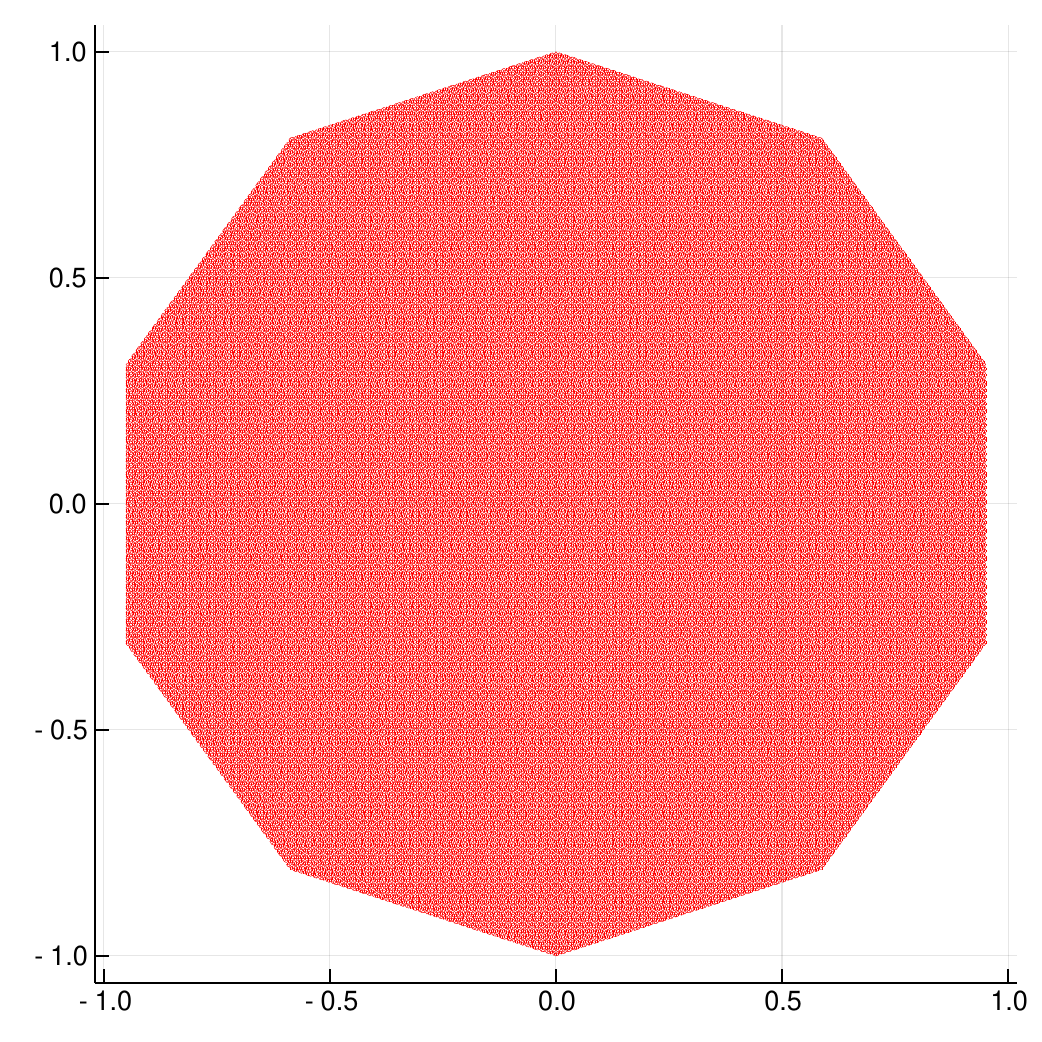}} 
        \end{tabular}
    \end{center}
    \caption{
    Penrose tilings with 143806 vertices.
    \label{fig:sfig1}
     }
    \end{figure}

\begin{figure}[t]
    \begin{center}
         \begin{tabular}{p{ 1 \columnwidth}} 
          \resizebox{1 \columnwidth}{!}{\includegraphics{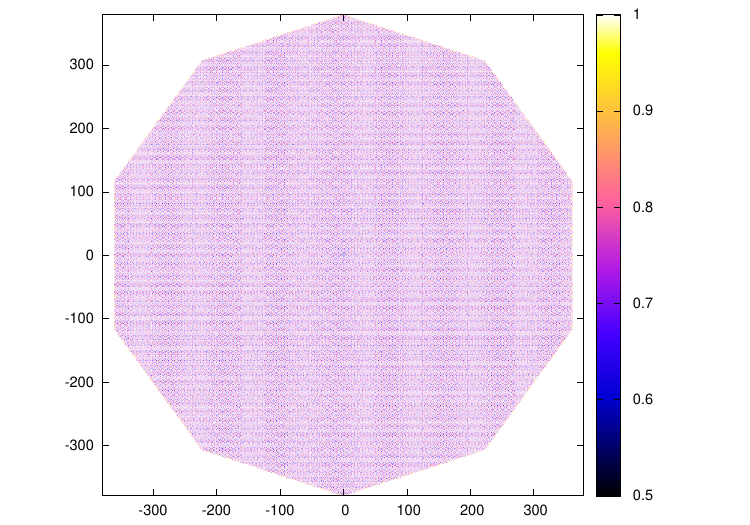}} 
        \end{tabular}
    \end{center}
    \caption{(Color online)
    Self-consistent solution of the Gap amplitude on the 143806-site Penrose lattice with a vortex located at a center.
    \label{fig:sfig2}
     }
\end{figure}

We show the Penrose lattice that we used in Fig.~\ref{fig:sfig1}. 
We regard each vertex of the rhombuses as a site, and put an electron hopping $t$ between two sites connected by the edge of the rhombuses. 
The Hamiltonian is expressed as 
\begin{align}
{\cal H} = -t \sum_{\langle ij \rangle \sigma} c_{i\sigma}^{\dagger} c_{j \sigma} - \mu \sum_{i \sigma} n_{i \sigma} + U \sum_i n_{i \uparrow} n_{i \downarrow},
\end{align}
where $n_{i \sigma} = c_{i\sigma}^{\dagger} c_{i \sigma}$. 
For simplicity, we neglect the Hartree mean-fields. 
The site-dependent superconducting order parameter is 
\begin{align}
\Delta_i = U \langle c_{i \uparrow} c_{i \downarrow} \rangle.
\end{align}
The self-consistent solution of the superconducting mean-fields is shown in Fig.~\ref{fig:sfig2}.

\begin{thebibliography}{99}
\bibitem{Nayak2008}
C. Nayak, S. H. Simon, A. Stern, M. Freedman, and S. Das Sarma, "Non-Abelian anyons and topological quantum computation", Rev. Mod. Phys. {\bf 80}, 1083 (2008).
\bibitem{Teo2010}
J. C. Y. Teo and C. L. Kane, "Majorana Fermions and Non-Abelian Statistics in Three Dimensions", Phys. Rev. Lett. {\bf 104}, 046401 (2010).
\bibitem{Alicea2011}
J. Alicea, Y. Oreg, G. Refael, F. von Oppen, and M. P. A. Fisher, "Non-Abelian statistics and topological quantum information processing in 1D wire networks", Nat. Phys. {\bf 7}, 412 (2011).
\bibitem{Wang2018} D. Wang, L. Kong, P. Fan, H. Chen, S. Zhu, W. Liu, L. Cao, Y. Sun, S. Du, J. Schneeloch, R. Zhong, G. Gu, L. Fu, H. Ding, and H. J. Gao, "Evidence for Majorana bound states in an iron-based superconductor", Science {\bf 362}, 333 (2018).
\bibitem{Machida2019}
T. Machida, Y. Sun, S. Pyon, S. Takeda, Y. Kohsaka, T. Hanaguri, T. Sasagawa, and T. Tamegai, "Zero-energy vortex bound state in the superconducting topological surface state of Fe(Se,Te)", Nat. Mater. {\bf 18}, 811 (2019).
\bibitem{Covaci2010}
L. Covaci, F. M. Peeters, and M. Berciu, "Efficient Numerical Approach to Inhomogeneous Superconductivity: The Chebyshev-Bogoliubov-de Gennes Method", Phys. Rev. Lett. {\bf 105}, 167006 (2010).
\bibitem{Nagai2012}
Y. Nagai, Y. Ota, and M. Machida, "Efficient numerical self-consistent mean-field approach for fermionic many-body systems by polynomial expansion on spectral density", J. Phys. Soc. Japan {\bf 81}, (2012).
\bibitem{Nagai2013} Y. Nagai, Y. Shinohara, Y. Futamura, Y. Ota, and T. Sakurai, "Numerical Construction of a Low-Energy Effective Hamiltonian in a Self-Consistent Bogoliubov-de Gennes Approach of Superconductivity", J. Phys. Soc. Japan {\bf 82}, 094701 (2013).
\bibitem{Nagai2017} Y. Nagai, Y. Shinohara, Y. Futamura, and T. Sakurai, "Reduced-Shifted Conjugate-Gradient Method for a Green’s Function: Efficient Numerical Approach in a Nano-Structured Superconductor", J. Phys. Soc. Japan {\bf 86}, 014708 (2017).
    \bibitem{Furukawa2004}
    N. Furukawa and Y. Motome, "Order N Monte Carlo Algorithm for Fermion Systems Coupled with Fluctuating Adiabatical Fields", J. Phys. Soc. Japan {\bf 73}, 1482 (2004).
    \bibitem{Dagotto1994}
   E. Dagotto, "Correlated electrons in high-temperature superconductors", Rev. Mod. Phys. 66, {\bf 763} (1994).
\bibitem{Ozaki2000}
T. Ozaki, M. Aoki, and D. G. Pettifor, "Block bond-order potential as a convergent moments-based method", Phys. Rev. B {\bf 61}, 7972 (2000).
\bibitem{Ozaki2006}
T. Ozaki, "{\it O(N)} Krylov-subspace method for large-scale {\it ab initio} electronic structure calculations", Phys. Rev. B {\bf 74}, 245101 (2006).
\bibitem{Kamiya2018}
 K. Kamiya, T. Takeuchi, N. Kabeya, N. Wada, T. Ishimasa, A. Ochiai, K. Deguchi, K. Imura, and N. K. Sato, "Discovery of superconductivity in quasicrystal", Nat. Commun. {\bf 9}, 154 (2018).
\bibitem{Sakai2017} S. Sakai, N. Takemori, A. Koga, and R. Arita, "Superconductivity on a quasiperiodic lattice: Extended-to-localized crossover of Cooper pairs", Phys. Rev. B {\bf 95}, 024509 (2017).
\bibitem{Sakai2019} S. Sakai and R. Arita, "Exotic pairing state in quasicrystalline superconductors under a magnetic fields", Phys. Rev. Res. {\bf 1}, 022002 (2019).
 \bibitem{sup} See Supplemental Material at [URL will be inserted by publisher for the detail for the Penrose lattice.]
  \bibitem{Hayashi1998}
 N. Hayashi, T. Isoshima, M. Ichioka, and K. Machida, "Low-Lying Quasiparticle Excitations around a Vortex Core in Quantum Limit", Phys. Rev. Lett. {\bf 80}, 2921 (1998).
  \bibitem{note} In the 375971-site Penrose lattice, the minimum energy level in a vortex core is about $0.25t$. 
 \bibitem{note2} 
Each node contains two Intel Xeon E5-2680v3 (2.5 GHz, 12 core) CPUs, giving a total of 60,240 compute cores, with 64 GB of RAM per node.

 \bibitem{Smith2016}
E. D. B. Smith, K. Tanaka, and Y. Nagai, "Manifestation of chirality in the vortex lattice in a two-dimensional topological superconductor", Phys. Rev. B {\bf 94}, 064515 (2016).
\bibitem{note3} The mixing ratio in the self-consistent calculation is $r = 0.05$ in this paper, where the next superconducting mean-fields are calculated by $\Delta^{\rm new}_i = r U \langle c_{i \uparrow} c_{i \downarrow} \rangle + (1-r) \Delta^{\rm old}_i$. 
 

\end{thebibliography}
\end{document}